\documentclass[aps, prl, amsmath, floats, floatfix, twocolumn,
superscriptaddress, nofootinbib]{revtex4-2}

\usepackage{graphicx}
\usepackage[export]{adjustbox}
\usepackage{caption}
\usepackage{color}
\usepackage{url}
\usepackage{bm}  
\usepackage{times}
\usepackage{amsmath}
\usepackage{amssymb}
\usepackage{float}
\usepackage[colorlinks=true,allcolors=blue]{hyperref}
\usepackage[normalem]{ulem}
\usepackage{hyperref}
\usepackage{todonotes}
\usepackage{subcaption} 
\usepackage[export]{adjustbox}
\usepackage{soul}

\begin{document}

\title{Nonlinear curvature effects in gravitational waves from inspiralling black hole binaries}
\author{Banafsheh Shiralilou}
\affiliation{GRAPPA, Anton Pannekoek Institute for Astronomy and Institute of High-Energy Physics, University of Amsterdam, Science Park 904, 1098 XH Amsterdam, The Netherlands}
\affiliation{Nikhef, Science Park 105, 1098 XG Amsterdam, The Netherlands}
\email{b.shiralilou@uva.nl}
\author{Tanja Hinderer}
\affiliation{GRAPPA, Anton Pannekoek Institute for Astronomy and Institute of High-Energy Physics, University of Amsterdam, Science Park 904, 1098 XH Amsterdam, The Netherlands}\affiliation{
Institute for Theoretical Physics,
Utrecht University, Princetonplein 5, 3584 CC Utrecht, The Netherlands}
\author{Samaya M. Nissanke}
\affiliation{GRAPPA, Anton Pannekoek Institute for Astronomy and Institute of High-Energy Physics, University of Amsterdam, Science Park 904, 1098 XH Amsterdam, The Netherlands}
\affiliation{Nikhef, Science Park 105, 1098 XG Amsterdam, The Netherlands}
\author{N\'estor Ortiz}
\affiliation{Instituto de Ciencias Nucleares (ICN), Universidad Nacional Autónoma de México (UNAM), Circuito Exterior C.U., A.P. 70-543, México D.F. 04510, México.}
\author{Helvi Witek}
\affiliation{Illinois  Center  for  Advanced  Studies  of  the  Universe \& Department of Physics, University of Illinois at Urbana-Champaign, Urbana, Illinois 61801, USA}

\begin{abstract}

Gravitational waves (GWs) from merging black holes allow for unprecedented probes of strong-field gravity. Testing gravity in this regime requires accurate predictions of gravitational waveform templates in viable extensions of General Relativity. We concentrate on scalar Gauss-Bonnet gravity, one of the most compelling classes of theories appearing as low-energy limit of quantum gravity paradigms, which introduces quadratic curvature corrections to gravity coupled to a scalar field and allows for black hole solutions with scalar-charge. 
Focusing on inspiralling black hole binaries, we compute the leading-order corrections due to curvature nonlinearities in the GW and scalar waveforms,
showing that the new contributions, beyond merely the effect of scalar field, appear at first post-Newtonian order in GWs.
We provide ready-to-implement GW polarizations and phasing. Computing the GW phasing in the Fourier domain, we perform a parameter-space study to quantify the detectability of deviations from General Relativity.
Our results lay important foundations for future precision tests of gravity with both parametrized and theory-specific searches.
\end{abstract}

\maketitle

\textbf{\textit{Introduction}}.
Gravitational waves (GWs) from merging black hole (BH) binaries are exploring new frontiers in strong-field gravity, \textit{e.g.,}~\cite{Abbott:2016blz}. A key challenge is to test whether Einstein's theory of General Relativity (GR) describes gravity accurately at all scales accessible to BHs, and to discover signatures of quantum gravity. Several BH mergers have already been detected by the LIGO and Virgo interferometers~\cite{LIGOScientific:2018mvr,Abbott:2020niy,Nitz:2018rgo,Venumadhav:2019lyq,Sachdev:2019vvd}. We anticipate an ever-increasing number and 
high precision measurements starting with the upcoming fourth observing run of the GW detector network.


To detect and to measure the properties of merging BH binaries, we crucially rely on detailed theoretical models to interpret GW signals. At present, models to test gravity are mostly null tests against GR, with parameterized deviations from GR waveforms. These tests are performed only on single coefficients~\cite{Yunes:2009ke,LIGOScientific:2019fpa,Abbott:2020jks} 
and thus, such strategies remain limited when interpreting theoretical constraints. Therefore, there is an urgent need to compute inspiral-merger-ringdown waveforms from alternative theories of gravity in order to allow for
an informed mapping of parametrized approaches to extensions of GR,
consistent, theory-specific
comparisons against observations, and for a systematic search of quantum gravity signatures in GW detections.

In this letter, we provide, for the first time, analytical  waveforms that include the effect of non-linear curvature corrections to inspiraling binaries beyond the weak-coupling limit,
for a well-motivated class of beyond GR theories. These theories contain contributions from the Gauss-Bonnet (GB) topological invariant class through the scalar $\mathcal{R}_{GB}=R^{2}-4 R^{\mu \nu} R_{\mu \nu}+R^{\mu \nu \rho \sigma} R_{\mu \nu \rho \sigma}$, which respects the Lorentz and CPT symmetries (this differs from the dynamical Chern–Simons theory~\cite{PhysRevD.68.104012,ALEXANDER20091}, for instance), and is coupled to a dynamical scalar. Scalar Gauss-Bonnet (sGB) gravity theories are ghost-free and arise in the low-energy limit of quantum gravity paradigms such as string theories and loop quantum gravity~\cite{Boulware:1985wk,GROSS198741,Kanti:1995vq}, which makes them promising effective theories at the energy scales relevant for astrophysical BHs. 
In sGB theories, BHs can spontaneously scalarize~\cite{Silva:2017uqg,Doneva:2017bvd} 
or develop scalar hair~\cite{Kanti:1995vq,Sotiriou:2014pfa,Benkel:2016rlz,Benkel:2016kcq,Ripley:2019irj,Antoniou:2017hxj,Antoniou:2017acq}. The scalar and higher curvature contributions modify BH binary's dynamics and GWs,
making BH mergers the most interesting avenue to test these theories.
 
Previous work on analytical models in 
sGB gravity has focused on the leading-order contributions to BH binary waveforms~\cite{Yagi:2011xp}, which are impacted only by the scalar field and not by the curvature nonlinearities, and on computing the Lagrangian for the dynamics~\cite{Julie:2019sab}.
The effects of an extra scalar field on GWs of binary inspirals have also been analytically computed in  scalar-tensor (ST) theories~\cite{Bernard:2018hta,Mirshekari:2013vb,Lang:2013fna,Lang:2014osa,Sennett:2016klh}, where, however, only neutron stars scalarize.

The first numerical relativity simulations of in sGB gravity used an effective-field-theoretical treatment~\cite{Witek:2018dmd,Okounkova:2020rqw} or the decoupling limit~\cite{Silva:2020omi} due to challenges in the time evolution formulation for general couplings~\cite{Julie:2020vov,Witek:2020uzz}. Recent progress in the formulation of sGB as well-posed initial value problem~\cite{Kovacs:2020ywu,Kovacs:2020pns} was used in the first numerical evolution of the nonlinear field equations~\cite{East:2020hgw}.

This work makes important progress on three fronts: (i) we compute, for the first time, analytical waveforms during the inspiral stage of binary evolution with the effect of higher curvature corrections.
Previous work in~\citep{Yagi:2011xp} captured only the corrections due to the scalar field; (ii) 
Our calculation and methods are not limited to the small-coupling approximation; they are applicable to all coupling strengths that lie within the theoretical bound~\cite{Kanti:1995vq,Sotiriou:2014pfa}
as well as general couplings that remain unconstrained;
(iii) We perform a parameter space study by varying the coupling parameter, coupling function, and BH masses,
for scalar as well as tensor radiation-dominated inspirals.
We further demonstrate that the effect of the GB scalar is distinct from the scalar in ST theories due to explicit GB coupling dependent terms. This has consequences for interpreting GW signals from BH-neutron star binaries~\cite{Carson:2019fxr}.

Using the Post-Newtonian (PN) approach, we compute the scalar and tensor waves to half and one relative PN-order ($\sim \mathcal{O}\,  ( 1/c )$ and $\sim \mathcal{O}\,  (1/c^2) $, where $c$ is the speed of light used here as the formal PN expansion parameter), respectively. We also calculate the GW phasing, to which measurements are very sensitive, as well as the polarization waveforms. Our results include higher order strong-field effects than previously computed, which are critical for GW measurements.
Such effects in alternative 
theories may mimic biases in fundamental source parameters when analysing with GR-only waveforms. This work lays the foundation for potential discoveries and provides the framework for computing new stringent constraints on nonlinear curvature effects of gravity.

\bigskip

\textbf{\textit{Black hole binaries in scalar Gauss-Bonnet theory}}. 
The gravitational action with the GB higher curvature terms is 
\begin{subequations}
\label{eq:totalaction}
\begin{equation}\label{action}
    S=\int d^4x\frac{c^3 \sqrt{-g}}{16\pi G}\left[R-2(\nabla \phi)^2+\alpha f(\phi)\, 
    \mathcal{R}_{GB}\right]\,.
\end{equation}
The coupling constant $\alpha$  has dimensions of length squared.
Choosing the coupling function $f(\phi)=e^{2\phi}/4$ corresponds to \textit{Einstein dilaton Gauss Bonnet} (EdGB) gravity~\cite{Kanti:1995vq}, and $f(\phi)=\phi$ to \textit{shift symmetric sGB} (ssGB) gravity~\cite{Sotiriou:2014pfa}.

The skeletonized matter action~\cite{1975ApJ...196L..59E} $S_m$  describing two BHs labeled by $A,B$, added linearly to $S$, is 
\begin{equation}\label{skeletonization}
    S_{m}
    =-c\int  M_{A}(\phi)\sqrt{-g_{\mu\nu}dx^{\mu}_{A}dx^{\nu}_{A}}+(A\leftrightarrow B).
\end{equation}
\end{subequations}
Here $x_{A}^{\mu}$ is the world line of particle A. With this ansatz, the self-gravity of the compact objects enters through the scalar dependent mass $M_{A}(\phi)$. In the weak-filed limit, it can be expanded as
\begin{equation}
    M_{A}(\phi)=m_{A}\Big[ 1+\alpha_{A}^{0}\delta\phi+\frac{1}{2}\left({\alpha^{0}_{A}}^{2}+\beta_{A}^{0}\right){\delta\phi}^2\Big] +\mathcal{O}(\delta\phi^3),
\end{equation}
with $\delta\phi$ the perturbation of $\phi$ around its background value $\phi_0$ and
$m_A$ the asymptotic value of the mass. The scalar-charge parameter and its derivative are defined as
\begin{equation}
    \left.\alpha_{A}^{0}=\frac{d\, ln\,M_{A}(\phi)}{d\phi}\right|_{\phi=\phi_{0}},
    \qquad 
    \left.\beta_{A}^{0}=\frac{d\, \alpha_{A}(\phi)}{d\phi}\right|_{\phi=\phi_{0}}.
\end{equation}

Within the small-coupling approximation, the explicit form of the scalar-charge $\alpha_{A}^{0}$ for nonspinning BHs has been derived to fourth order in the coupling in \citep{Julie:2019sab}. To the leading order, $\alpha_{A}^{0}\equiv -\alpha f'(\phi_{0})/2m_{A}^2$ for all coupling functions.
\bigskip

\textbf{\textit{Gravitational and scalar radiation}}.
To compute the dynamics and gravitational radiation of BH binaries in the theory~\eqref{eq:totalaction}, we introduce the {\it gothic metric} $\mathfrak{g}^{\alpha\beta }=\sqrt{-g} g^{\alpha\beta}$ and decompose it in the weak-field limit as $\mathfrak{g}^{\mu\nu}=h^{\mu\nu} + \eta^{\mu\nu}$, where $\eta_{\mu\nu}$ is the flat metric and $h_{\mu\nu}$ is the tensor perturbation.
Specializing to harmonic gauge, where $\partial_{\nu}\mathfrak{g}^{\mu\nu}=0$,
we write the field equations derived from action \eqref{eq:totalaction} in a \textit{relaxed} form~\cite{1975ctf..book.....L}, finding that
\begin{subequations}
\label{eq:eoms}
\begin{align}\label{wave}
   & \Box h^{\alpha\beta}=\frac{16\pi G}{c^4} (-g)T^{\alpha\beta}_{m}+\Lambda_{GB}^{\alpha\beta}+\Lambda^{\alpha\beta}_{GR}\,,\\
  &  \Box \phi=\frac{4\pi G}{c^4}\frac{S_{m,\phi}}{\sqrt{-g}}-\frac{\alpha f'(\phi)}{4} 
  \mathcal{R}_{GB}\,,
    \end{align}
    \end{subequations}
where $\Lambda^{\alpha\beta}_{GR}$ contains terms that are quadratic in $h^{\alpha\beta}$ and its derivatives
~\cite{1975ctf..book.....L} and $T^{\alpha\beta}_{m}$ is the stress-energy tensor, derived from the matter action in Eq.~\eqref{skeletonization}.
For the explicit GB contribution to the metric field equation we find
   \begin{equation}
   \begin{split}
 \Lambda_{GB}^{\alpha\beta}=
 &-8\alpha (-g)\, ^{*}\hat{R}^{*c\alpha\beta d}\nabla_{cd}f(\phi)\\
&+ \nabla_{c}\phi\nabla_{d}\phi\big(2\mathfrak{g}^{\alpha c}\mathfrak{g}^{\beta d}-\mathfrak{g}^{\alpha \beta}\mathfrak{g}^{c d}\big)\,,
 \end{split}
\end{equation}
where, $^{*}\hat{R}^{*c\alpha\beta d}$ is the gauge-fixed dual Riemann tensor.
The formal solutions to Eq.~\eqref{eq:eoms} are
computed with the retarded Green's function approach
\begin{equation}
h^{\mu \nu}(t, \mathbf{x})=\frac{1}{4\pi}\int \frac{s^{\mu \nu}\left(t^{\prime}, \mathbf{x}^{\prime}\right) \delta \left(t^{\prime}-t+\left|\mathbf{x}-\mathbf{x}^{\prime}\right|/c^2\right)}{\left|\mathbf{x}-\mathbf{x}^{\prime}\right|} d^{4} x^{\prime},
\label{Twaveall}\end{equation} where $s^{\mu\nu}$ denotes the source terms on the right hand side of \eqref{wave}, and similarly for the scalar field. 
The integral in Eq.~\eqref{Twaveall} extends over the past light cone of the field point $(ct,\mathbf{x})$. 
 To calculate the solution of the integral, we split the spacetime into three regions: (i) The strong-field zones close to each of the BHs. At the boundaries of these zones we extract the masses $M_A(\phi)$, and treat their interior regions as a skeletonized worldline~\cite{Dixon:1970zz,1975ApJ...196L..59E}; 
 (ii) The near-zone, where the separation between source and field point is less than the characteristic wavelength of the GWs; and (iii) The far zone at larger distances. With this splitting, we can use the post-Newtonian Direct Integration of the Relaxed Einstein equations (DIRE)~\cite{Will:1996zj} formalism to divide the integration of Eq.~\eqref{Twaveall} into four different calculations, corresponding to the near and far zone contributions for each relative location of the source and field points. 

More specifically, to turn the formal solutions of  Eq.~\eqref{Twaveall} into a practical scheme,
we make the PN assumption that $h_{\mu\nu}$ and $\delta\phi$ are small, and perturbatively expand the nonlinear terms in
$s^{\mu\nu}$ and its scalar analogue using the formal expansion parameter $1/c^2$,
keeping terms up to the relative first PN order. 
We follow the methods of~\cite{Will:1996zj} for evaluating the four different contributions to the integrals, and compute the equations of motion from the 1PN Lagrangian given in Ref.~\cite{Julie:2019sab} to eliminate accelerations. The technical details of the calculations are given in \cite{inprep}.

The energy radiated in tensor (T) and scalar (S) waves is computed from
\begin{equation}\dot{E}=\dot{E}_T+\dot{E}_S=\frac{c^3R^{2}}{32 \pi G} \oint \left[\dot{h}_{\mathrm{TT}}^{i j} \dot{h}_{\mathrm{TT}}^{i j} + \dot{\phi}^{2}\right] d^{2} \Omega\,,
\end{equation}
where $R$ is the distance between the source and the far zone field point \textit{i.e.}, the detectors, and $\mathrm{TT}$ denotes the transverse-traceless projection. In the following, we specialize to circular-orbit binary systems. In analogous to ST theories~\cite{Damour:1992we}, we define the binary parameters
\begin{eqnarray}\label{EOMparams}
&&\bar{\alpha} \equiv \left(1+\alpha_{A}^{0} \alpha_{B}^{0}\right)\,, \;\; \; \!\!
\bar{\gamma}\equiv-2 \frac{\alpha_{A}^{0} \alpha_{B}^{0}}{\bar{\alpha}}\,,\; \; \;\!\!
\bar{\beta}_{A} \equiv \frac{1}{2} \frac{\beta_{A}^{0} (\alpha_{B}^{0})^{ 2}}{\bar{\alpha}^{2}}\,,\nonumber\\
\end{eqnarray}
where $\Delta m=m_A-m_B$, with the convention $m_A<m_B$, and $m=m_A+m_B$ is the total mass.
We obtain, omitting corrections of $O(c^{-4})$ :
\begin{eqnarray}
&\dot{E}_T&=\bar{\cal F}^{\rm N}\left\{1+
{\cal F}^{\rm 1PN}_{\rm GR}
-\frac{16\tilde\beta_{+}\bar{v}^2}{3c^2} -\frac{10\bar{\gamma}\bar{v}^2}{3c^2}-\frac{\epsilon f'(\phi_{0})\bar{v}^6}{G^2\bar{\alpha}^{5/2}c^2}
\right.\nonumber\\
&&\left. \left[8\mathcal{S}_{1,1,0}+\frac{
16{\cal S}_{3,1,0}}{3\bar{\alpha}}
-\frac{261(
 \mathcal{S}_{1,0,1}-2 \eta\mathcal{S}_{1,0,-1})}{7}\right]\right\},\\
  &\dot E_S&=\bar{\cal F}^{\rm D}\left[\mathcal{S}^{2}_{-}-\frac{2 {\cal S}_-^2(20\tilde{\beta}_{+}+5 \bar{\gamma}-2\eta)}{ 3}\frac{\bar{v}^2}{c^2}\right.\nonumber\\
   &&+\frac{(4{\cal S}_+^2 -54{\cal S}_{-}^2) }{5 }\frac{\bar{v}^2}{c^2}-\frac{8 {\cal S}_{-}}{\bar{\gamma}}\left(\mathcal{S}_{-}\tilde{\beta}_{+}+\mathcal{S}_{+}\tilde{\beta}_{-}\right)\frac{\bar{v}^2}{c^2}\qquad\,\nonumber\\
  &&\left.\hspace{-2mm}-\frac{\bar{v}^6}{c^2}\frac{ \epsilon f'(\phi_{0})}{G^2\bar{\alpha}^{5/2}}\left(\frac{32{\cal S}_{3,1,0}\mathcal{S}^{2}_{-}}{3\bar{\alpha}}+\frac{\eta\Delta m{\cal S}_+\mathcal{S}_-{\cal S}_{1,1,0}}{8 m }\right)\right].\qquad \label{eq:EdotS}
\end{eqnarray}
Here, $\eta=m_Am_B/m^2$ is the symmetric mass ratio. Note that the circular-orbit velocity $\bar{v}=(Gm\bar\alpha\omega)^{1/3}$ differs from its GR definition by a factor of $\bar{\alpha}$.
The leading order energy flux in tensor radiation is $\bar{{\cal F}}^{\rm N}=32\eta^2 \bar{v}^{10}/(5G c^5\bar{\alpha}^2)$, where $N$ denotes the Newtonian order contribution. The contribution up to 1PN order ${\cal F}^{\rm 1PN}_{\rm GR}$ is given e.g. in~\cite{Blanchet:2013haa}. The prefactor of the leading order flux of scalar radiation due to dipole emission is $\bar{\cal F}^{\rm D}= 4\eta^2\bar{v}^8/(3G\bar{\alpha} c^3 )$. We have also defined 
\begin{eqnarray}\label{eq:AGBdef}
&&\mathcal{S}_{\pm}=\frac{\alpha_A^0\pm\alpha_B^0}{2\sqrt{\bar{\alpha}}},\quad  
\tilde{\beta}_{\pm}=\frac{\bar{\beta}_{A}(1-\frac{\Delta m}{m})\pm\bar{\beta}_{B}(1+\frac{\Delta m}{m})}{2},\nonumber\\
&&\epsilon=\frac{ \alpha}{ m^2}\,,\qquad
 {\cal S}_{a,b,c}=a {\cal S}_++\left(b\frac{\Delta m}{m}+c\right){\cal S}_-\,,
\end{eqnarray}
where $\epsilon$ is the dimensionless coupling parameter.

our result for the energy fluxes have a similar structure to those of a ST theory (see~\cite{Lang:2013fna,Lang:2014osa} ) but differ through the additional $\epsilon$-dependent terms entering first at relative 1PN order.This feature can be used in distinguishing the two theories when analysing BH-neutron star binaries.
We also note that the scaling of the GB contributions in the PN expansion is ${\cal O}(c^{-2})$,
irrespective of the value of the coupling. However, due to the different scaling with $\bar{v}$, the GB contributions ($\sim \bar{v}^6$) are suppressed at large separation compared to the other 1PN terms ($\sim \bar{v}^2$).

From the Lagrangian in~\cite{Julie:2019sab} we derive the circular-orbit binding energy~\cite{inprep} to ${\cal O}(c^{-2})$:
\begin{equation}\label{bindingenergy_r}
E=-\eta m \bar{v}^2\left[E^{1\text{PN}}_{\text{GR}}+\frac{(2\tilde\beta_{+}-\bar\gamma)\bar{v}^2}{3c^2}+\frac{11 \mathcal{S}_{3,1,0}\bar{v}^6}{3\bar{\alpha}c^2}\frac{ \epsilon f'(\phi_{0})}{G^2\bar{\alpha}^{5/2}}\right],
\end{equation}
where $E^{1\text{PN}}_{\text{GR}}$ is the 1PN correction in GR~\cite{Blanchet:2013haa}.

\bigskip

\textbf{\textit{Gravitational wave phasing}.} The GW measurements are very sensitive to the phase evolution of the waveform.
An approximation for the phasing can be derived from energy balance 
$dE(\bar{v})/dt=-\mathcal{F}(\bar{v})$, which is valid as long as $\dot{\omega}/\omega^2\ll 1$. This yields the 
differential equations
\begin{equation}\label{orbphase_t}
\begin{aligned}
\frac{d \varphi}{d t}-\frac{ \bar{v}^{3}}{G\bar{\alpha}m} &=0\,,\qquad
\frac{d \bar{v}}{d t}+\frac{\mathcal{F}(\bar{v})}{ E^{\prime}(\bar{v})} &=0.
\end{aligned}
\end{equation}
We solve this system in the \textit{Taylor T4} approximation~\cite{Buonanno:2009zt}, by expanding the entire ratio $\mathcal{F}(\bar{v})/E'(\bar{v})$ to 1PN order and solving Eq.~\eqref{orbphase_t} numerically for the phase evolution.

As the parametrized tests of gravity are mainly based on waveforms in the Fourier domain, we also compute the Fourier domain phase $\psi(f)$ at the dominant GW frequency $f=\omega/\pi$ in the stationary phase approximation (SPA) ~\citep{PhysRevD.59.124016} by using
\begin{equation}
\label{eq:FourierphaseSPA}
    \psi(f)=2\pi f t_{0}-\phi_{0}+2\int_{v_f}^{v_0}(v_f^3-v^3)\frac{E'(v)}{\mathcal{F}(v)}dv\,.
\end{equation}
The subscript $0$ refers to a reference point in the binary evolution. To solve for the GW phase from \eqref{eq:FourierphaseSPA}, we distinguish systems whose inspiral is driven by scalar-dipolar versus tensor-quadrupolar radiation, with the scalar-dipolar driven (DD) regime relevant for:
\begin{equation}\label{DDtoQD}
\bar{v}_{\rm DD}^2\ll \frac{5c^2{\cal S}_-^2\bar\alpha}{24}
\;\; {\rm or} \;\; f^{\rm DD}\ll \left(\frac{5}{24}\right)^{3/2} \frac{c^3\, {\cal S}_-^3 \, \sqrt{\bar\alpha}}{\pi G m}.
\end{equation}
At higher frequencies the system is quadrupole driven (QD).
The phase evolution in the QD regime for equal masses (mass ratio $q=m_A/m_B=1$) is given by,
\begin{eqnarray}
    \psi^{\rm QD}_{q=1}&=&\frac{3 c^5\bar{\alpha}}{128 \bar{v}^5\bar{\xi}}\left\{1+\frac{20\bar{v}^2}{9c^2}\left[\frac{1247}{336\bar{\xi}}-\frac{3}{2}+\left(\frac{980}{336\bar{\xi}}-\frac{1}{6}\right)\eta\right.\right.\nonumber\\
    &&\left.\hspace{-8mm}+\left(\frac{448}{336\bar{\xi}}-\frac{4}{3}\right)\bar{\gamma}+2\left(\frac{896}{336\bar{\xi}}+\frac{4}{3}\right)\tilde{\beta}_{+}\right]
    -\frac{25c^{\rm 1PN}_{\rm S}\bar{\alpha}\bar{v}^2}{54 \bar{\xi} c^2}\nonumber\\
    &&\hspace{-8mm}\left.-\frac{40\mathcal{S}_{+}\bar{v}^6}{c^2}\frac{f'(\phi_{0})\epsilon}{g^2 \bar{\alpha}^{5/2}} \Bigg(\frac{12}{\bar{\xi}}+\frac{495(1-2\eta)}{28\bar{\xi}}
    +\frac{88}{\bar{\alpha}} \Bigg)\right\},\qquad
\end{eqnarray}
with $\bar{\xi}=1+{\cal S}_+^2\bar{\alpha}/6$.
The full expression for arbitrary masses and for the DD regime is given in Ref.~\cite{inprep}.
We note that the QD phasing has contributions from 1PN scalar flux, indicated as $c_{S}^{\text{1PN}}$. We calculated the scalar waveform to 0.5PN order, leaving this contribution undetermined. Following the strategy employed for 2PN tidal effects~\cite{Damour:2012yf}, we will keep all the other 1PN terms and set the missing contributions to zero. The $c_{S}^{\text{1PN}}$ term is expected to have a similar structure as other terms and possibly depend on the parameter $\mathcal{S}_{+}$. The inclusion of this term would increase the overall energy flux and thus the phase differences between sGB and GR.

\bigskip

\textbf{\textit{Ready-to-use gravitational wave polarizations}.}
\begin{figure*}[hbt!]
\centering
\includegraphics[width=\textwidth]{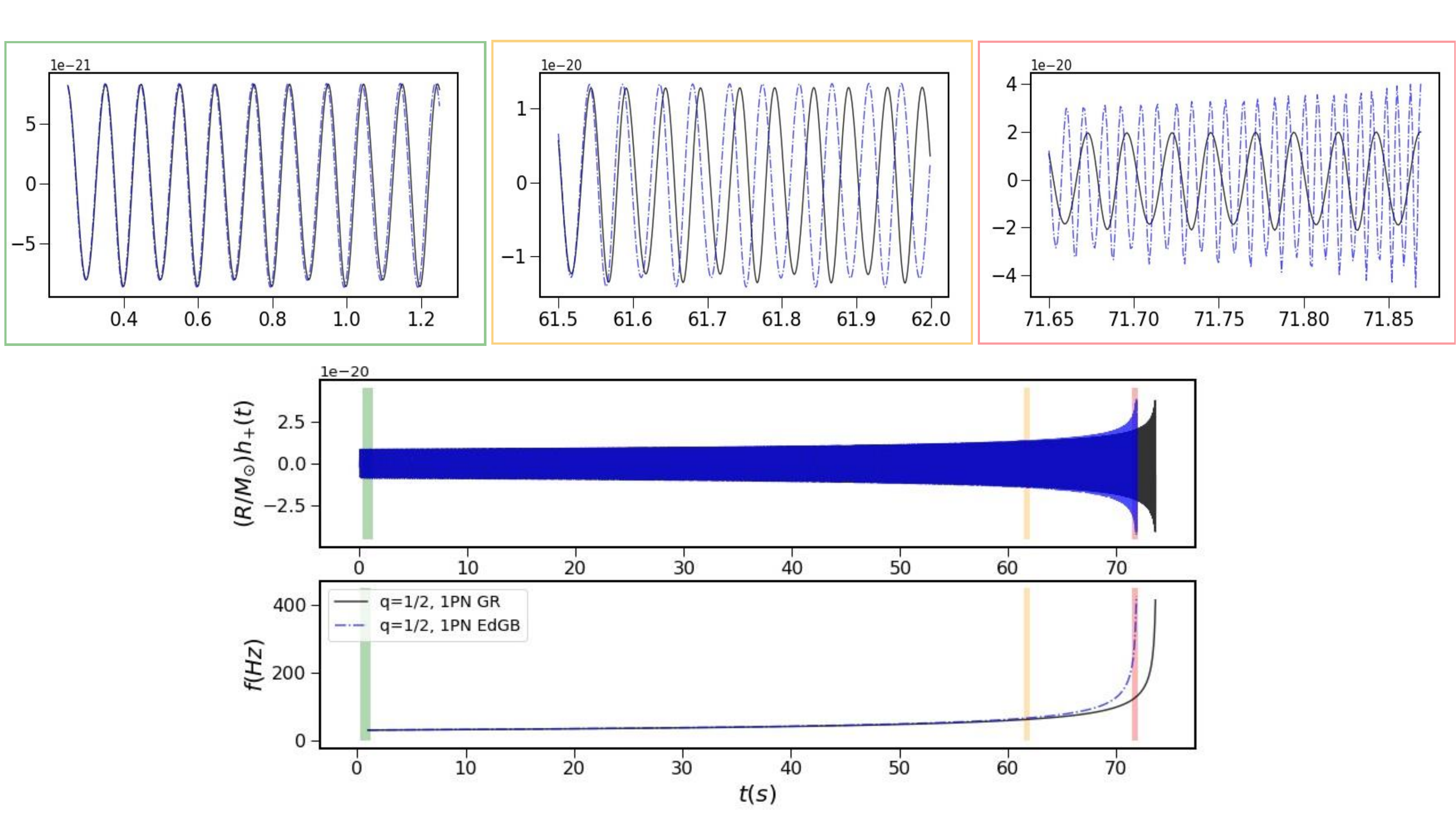}
\caption{The time evolution of GW signal for a $m=15M_{\odot}$ binary with $q=1/2$ and $\epsilon=0.03$, and the corresponding GW frequency evolution. The blue dashed curve indicates the EdGB waveform and the black curve the 1PN GR waveform. The orbit is viewed edge-on ($i=\pi/2$) and $t=0$ indicates the time corresponding to $f=10 Hz$. The shaded colored regions correspond to different snapshots of the waveform.}
\label{waveform}
\end{figure*}
In the time domain, GW detectors measure the linear combination of polarization waveforms $h_{+}(t)$ and $h_{\times}(t)$.
We derive the two GW polarizations from the solution to Eq.~\eqref{Twaveall}, solved explicitly in~\citep{inprep}. 
To 1PN order, we obtain
\begin{eqnarray}
    h_{+,\times}&=&\frac{2G\mu}{Rc^2}\frac{\bar{v}^2}{c^2}\left[H^{0}_{+,\times}+\frac{\bar{v}}{c} H^{1/2}_{+,\times}+\frac{\bar{v}^2}{c^2} H^{1}_{+,\times}\right.\nonumber\\
    &&\quad\quad \left.+\frac{\bar{v}^6}{c^2}\frac{\epsilon f'(\phi_{0})}{24G^2\bar{\alpha}^{5/2}} H^{1,\rm GB}_{+,\times} +{\cal O}(c^{-3})\right].
\end{eqnarray}
where the normal to the orbit differs from the radial direction to the observer by an inclination angle $\i$.
The coefficients of the plus polarization are
\begin{equation}
    \begin{split}
    &H^{0}_{+}=H^{0}_{+(GR)}\,,\qquad\qquad\qquad\qquad\quad H^{1/2}_{+}=H^{1/2}_{+(GR)}\,,\\
    &H^{1}_{+}=H^{1(GR)}_{+}
    +\frac{2}{3}(\bar{\gamma}+\bar{\beta}_{+})\left(1+\cos ^{2} (i)\right) \cos (2 \varphi)\,,\\
    &H^{1}_{+(GB)}=
     192\left[(\cos (2i)+3) \cos (2\varphi) \mathcal{S}_{2,1,0}+\right.\\
    &\left.\,\sin ^2(i) \mathcal{S}_{3,1,0}
    \right]+32 \mathcal{S}_{3,1,0} \left[\left(\cos ^2(i)+1\right) \cos (2\varphi)-3 \sin ^2(i)\right]\\
    &\, +18 \big[
    (2\eta +1) \mathcal{S}_{-}+(1-2 \eta) \mathcal{S}_{+}
    \big]\\
    &\,\left[2 \sin ^2(2i) \cos (2\varphi)-\sin ^2(i) (\cos (2 i)+3) (3 \cos (4 \varphi)+1)\right]\,,
    \end{split}
\end{equation}
And for the cross polarization, they are
\begin{equation}
    \begin{split}
    &H^{0}_{\times}=H^{0(GR)}_{\times}\,,\qquad\qquad\qquad\qquad\quad
    H^{1/2}_{\times}= H^{1/2(GR)}_{\times} \,,\\
    &H^{1}_{\times}=H^{1(GR)}_{\times}+\frac{4}{3}\cos(i)\sin(2\varphi)\Big[\bar{\gamma}+2\beta_{+}-2\frac{\Delta m}{m}\beta_{-}\Big]\,,\\
   & H^{1}_{\times(GB)}=
    \cos (i)\Big\{2 \sin (2\varphi) \\
   &\,\left[9 \sin ^2(i) (\mathcal{S}_{108,52,0}-3+(2 \eta+1)\mathcal{S}_{-}+(1-2 \eta)\mathcal{S}_{+})\right]\\
   &\,-27 \sin ^2(i) \sin (4\varphi) \big[(2 \eta+1)\mathcal{S}_{-}+(1-2 \eta) \mathcal{S}_{+}\big]\Big\}.
    \end{split}
\end{equation}

In FIG.\ref{waveform} we show the GW polarization and phase evolution in time.
Exemplarily, we choose an intermediate coupling value $\epsilon=0.03$ for a BH binary with $m=15M_{\odot}$ and $q=1/2$. We also show a comparison to the 1PN waveforms within GR. The evolution starts at a GW frequency of $f=10Hz$, i.e., when the GWs would first enter the sensitivity band of current ground-based GW detectors.
We observe that the dephasing of the waves starts
early in the evolution, while the difference between the amplitudes remains relatively small until the binary reaches frequencies of around $f\approx 60Hz$, after which the GB phasing and amplitude increase rapidly and the difference with GR waveform becomes significant.

\bigskip

\textbf{\textit{Impact of  higher-curvature gravity on GWs}}. 
\begin{figure*}[hbt!]
\centering
\includegraphics[width=1.07\linewidth]{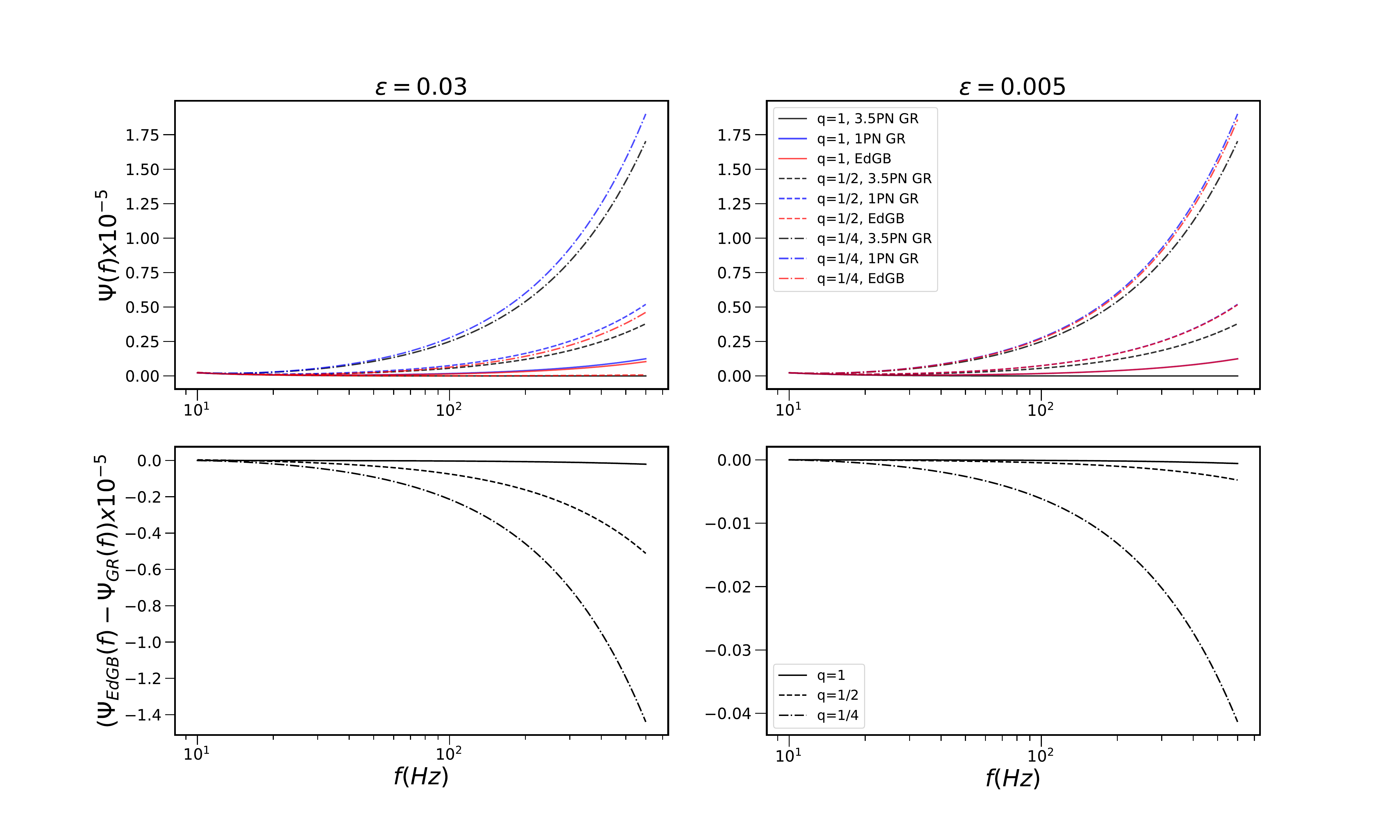}
\caption{Top: GW inspiral phase $\psi(f)$ as a function of frequency $f$ for a $m=15 M_{\odot}$ binary
with $q=1$ (solid lines), $q=1/2$ (dashed lines), and $q=1/4$ (dot-dashed lines), with $\epsilon=0.03$ (left) and $\epsilon=0.005$ (right). The red curves indicate EdGB gravity. In blue, we show the corresponding 1PN GR phase, and in black, the 3.5PN GR phase. Bottom: GW phase difference between EdGB $\psi_{\text{EdGB}}$ and 1PN GR $\psi_{\text{GR}}$ shown for aforementioned systems. }
\label{phaseevol}
\end{figure*}
Considering EdGB and ssGB theories, we study the impact of the GB coupling parameter on the phase evolution of quasi-circular BH binaries in the A+LIGO sensitive frequency band~\cite{TheLIGOScientific:2014jea}. 

Requiring regular BH horizons limits the coupling parameter to $\epsilon<0.619$ in EdGB~\cite{Kanti:1995vq} and $\epsilon\lesssim 0.3$ in ssGB~\cite{Sotiriou:2014pfa}. 
Simulations of BH mergers predict a bound on the coupling of $\epsilon\lesssim0.03$ (i.e. $\sqrt{\alpha}\lesssim3.2$~km for an equal-mass binary of $20M_{\odot}$) from current GW observations~\cite{Witek:2018dmd,Okounkova:2020rqw}. This is consistent with the GW-based constraints of~\cite{Nair:2019iur} and constraints from low-mass X-ray binaries~\cite{Yagi:2012gp}. 

Here, we choose $\epsilon= 0.03$, in correspondence with Fig.~\ref{waveform}, and also $\epsilon=0.005$. As we will show, the latter choice marks the threshold for detectability of sGB phase modifications, for many low-mass BH binary systems. We use the explicit result of~\cite{Julie:2019sab} for the scalar charges, valid to first order in the coupling.
As we are mainly interested in the behaviour of the theory at high curvature regimes (\textit{i.e.} low BH masses), we choose a total mass of $m=15M_\odot$ and vary the individual BH masses for mass ratios $q=1, 1/2, 1/4$. For the binaries with $q=1$, the scalar radiation is very small, as $\mathcal{S}_{-}$ vanishes in this case. 

The GB corrections to the inspiral phase evolution are determined by the GB coupling parameter $\alpha=\epsilon m^2$, which also sets the scalar charges. 
For instance, the threshold \eqref{DDtoQD} indicates that for a relatively large $\epsilon$, those systems with $1/2\leq q<1$ are DD when they enter the LIGO band.
Yet for small $\epsilon$, having a DD regime and transition to QD in the ground-based detector bands requires low mass ratio binaries with individual BHs as light as $2M_{\odot}$. This means that for such small couplings, binary BHs are typically QD systems in the ground-based detector bands, yet mixed binaries may have a DD regime. 
For example, in the case of $15M_{\odot}$ BH binaries with $\epsilon=0.1$, the $q=1/2$ system is a DD inspiral that transitions to the QD regime, and the $q=1/4$ case is DD throughout the entire inspiral.

In Fig.\ref{phaseevol}, we show the phase evolution of binary BHs in EdGB gravity as compared to the corresponding phase in GR to 3.5PN order, for the aforementioned choices of $\epsilon$. To isolate the GB effects, we also compare the phasing with that of GR to 1PN order.
The upper frequency bound is chosen as $f_{\text{max}}=2(6^{3/2}\pi m)^{-1}\approx586\,\mathrm{Hz}$ and to simplify the comparison, all phases are aligned with the 1PN equal-mass phase in GR at the minimum frequency limit.
These systems represent the most relevant regime for the majority of binaries observable with the current detectors LIGO/Virgo/KAGRA.

We only show the EdGB phase evolution
as the phase difference between the ssGB and EdGB theories is relatively small compared to the overall phase evolution. This is to be expected as we are using a first order approximation to 
$\alpha_{A}^{0}$. We note here that for $q\neq1$ binaries, this difference is within the limit of detectability once
having $\epsilon > 0.1$ ,\textit{i.e.}, for $\epsilon=0.01$ the phases differ by $\mathcal{O}(10)$ GW cycles.

As shown in Fig.~\ref{phaseevol}, the sGB phases are always less than their 1PN GR analogue, decreasing the overall phase of an equal-mass BH binary by
$\sim 322$ GW cycles if $\epsilon=0.03$, and by
$\sim 9$ GW cycles if $\epsilon=0.005$. As can be seen from the plot, this phase difference increases significantly for $q\neq 1$ binaries, which also emit energy through scalar dipole radiation.
Overall, decreasing the value of $\epsilon$ results in smaller deviations from the GR phase. For very small values of the coupling parameter (not shown here) such as $\epsilon=0.001$, the change in number of GW cycles of binaries with $q<1/2$ is 
of the order of several cycles, making the GB effects still within the limit of detectability.

\bigskip

\textbf{\textit{Conclusions}}.
We have studied GWs from BH binary inspirals for gravity theories with higher-curvature corrections characterized by the coupling of the GB invariant to a scalar field. We have computed novel signatures from nonlinear curvature corrections to 1PN order beyond the leading quadrupole emission in the gravitational waveform, and to 0.5PN order in the scalar waveform, in addition to scalar effects considered in previous work~\cite{Yagi:2011xp}.
We provide ready-to-implement 1PN inspiral GW templates.
By deriving the SPA gravitational phase and evaluating it for examples of BH binaries in ssGB and EdGB theories, we show that the inspirals are accelerated compared to
1PN GR case, with the deviation being strongly dependent on the coupling parameter of the theory.

Our results are not restricted to specific choices of the coupling function nor to the weak coupling limit.
In particular, they allow to investigate a wide class of sGB gravity
including those that yield spontaneously scalarized BHs~\cite{Silva:2017uqg,Doneva:2017bvd}, a truly nonlinear effect that is suppressed by a weak coupling treatment.
Thus, our work lays the foundation to explore dynamical scalarization or descalarization of BH binaries~\cite{Silva:2020omi} during the early inspiral.


Our results provide a critical first step towards constructing inspiral-merger-ringdown GW templates at high curvature regimes
and provides a useful benchmark for numerical relativity simulations of the merger phase~\citep{inprep}.
By further showing that the scalar-charge induced dipole radiation as well as the higher curvature effects are potentially observable in A+LIGO/Virgo/KAGRA sensitivity bands, we provide the baseline for more extensive parameter estimation studies, which we leave for future work, for both ground-based and multi-band GW observations.

\textbf{\emph{Acknowledgments.}} We thank L.~Gualtieri, H.~O.~Silva, H.~S.~Chia and  S.~Mukherjee for useful discussions. BS, TH and SMN are grateful for financial support from the Nederlandse Organisatie voor Wetenschappelijk Onderzoek (NWO) through the Projectruimte and VIDI grants (Nissanke). TH also acknowledges financial support from the NWO sectorplan. NO acknowledges financial support by the CONACyT grants ``Ciencia de frontera" 140630 and 376127, and by the UNAM-PAPIIT grant IA100721.HW acknowledges financial support provided by NSF Grant No. OAC-2004879 and the Royal Society Research Grant No. RGF\textbackslash R1\textbackslash 180073.




%

\end{document}